\documentstyle[12pt]{article}
\begin{document}
\title{INTEGRABLE KdV SYSTEMS:\\Recursion Operators of Degree Four}
\author{ Metin G{\" u}rses\\
{\small Department of Mathematics, Faculty of Sciences}\\
{\small Bilkent University, 06533 Ankara - Turkey}\\
{\small and}\\
Atalay Karasu\\
{\small Department of Physics , Faculty of Arts and  Sciences}\\
{\small Middle East Technical University , 06531 Ankara-Turkey}}
\date{}
\begin{titlepage}
\maketitle


The recursion operator and bi-Hamiltonian
formulation of the Drinfeld-Sokolov system are given.

\end{titlepage}

Recently \cite{MET}, \cite{MET1} we have given a subclass
of the coupled system of Korteweg -de Vries (KdV) equations. These
system of equations are mainly classified ,if they are integrable ,
with respect to a pair of numbers ($m$, $n$). Here the numbers $m$
and $n$ are respectively the highest powers of the operators $D$ and
$D^{-1}$ in the recursion operator ,${\cal R}$.  The type (2,1)
has been extensively studied by several authors
\cite{MET}-\cite{FOK}. The Svinolupov Jordan
KdV system \cite{SVI}, \cite{SV1} correspond to the type (2,2) and
some mixed cases were considered quite recently \cite{OLV2}. In this work we
start to consider the type (4,1) for $N=2$. Our preliminary
classification includes the systems: Hirota-Satsuma \cite{HIR},\cite{FUC}
, \cite{DOD}-\cite{XIA}, Boussinesq \cite{MC}-\cite{OLV},
and Drinfeld-Sokolov (DS) systems \cite{DRI1},\cite{DRI}.
 The first two systems are
well known to be integrable, that is they admit a recursion operator and
also a bi-Hamiltonian structure.
The latter one admits a Lax-pair \cite{DRI1},\cite{BOG}
and its  Painlave truncated Backlund transformations are studied by Tian
and Gao \cite{TIA}.

Here we show that DS system admits a hereditary
recursion operator and hence results out of a Hamiltonian pair.
We consider a system of $N$ nonlinear equations which is called a coupled
KdV system
\begin{equation}
 q^{i}_{t}=b^{i}_{j}q^{j}_{xxx}+s^{i}_{j k}q^{j}q^{k}_{x}+\chi^{i}_{j}
q^{j}_{x},\label{a0}
\end{equation}

\noindent
where $i,j,k=1,2,...,N$ ,
$q^{i}$ are functions depending on the variables $x$ , ${t}$\, , \,
and  $b^{i}_{j}$\,\, ,\,\,$s^{i}_{jk}$ and $\chi^{i}_{j}$
are constants. Here we use the Einstein convention,i.e.,repeated
indices are summed up over $1-N$. The part containing the terms
($b^{i}_{j}, s^{i}_{jk}$) will be called the principle part of
the system of KdV equations (\ref{a0}). In the classification of the
above system (\ref{a0}) singular and non-singular behavior of
the matrix $b^{i}_{j}$ plays an essential role
\cite{MET1}. The system is called degenerate if $b$ is singular ,i.e.,
$ det(b^{i}_{j})=0$ otherwise it is called non-degenerate. The Hirota-
Satsuma system is an example of a non-degenerate and the Boussinesq
system is an example of a degenerate cases.

We propose that the recursion operator (if it exists) of the system of
equations in (\ref{a0}) takes the form

\begin{eqnarray}
{\cal R}^{i}_{j} & =& a^{i}_{j}\,D^{4}+D^{i}_{jk}\,q^{k}D^{2}+t^{i}_{j}\,D^{2}
+\xi^{i}_{j}D+c^{i}_{jk}\,q^{k}_{x}\,D
+ R^{i}_{jk}\,q^{k}_{xx}+S^{i}_{jlm}q^{l} q^{m} \nonumber\\
& & +Q^{i}_{jk}\,q^{k}
 +b^{i}_{jk}\,q^{k}_{xxx}\,D^{-1} +M^{i}_{jlm}q^{l} q^{m}_{x}D^{-1}
+N^{i}_{jlm}q^{l}_{x}D^{-1}q^{m} \nonumber\\
& & +P^{i}_{jk}q^{k}_{x}D^{-1}+w^{i}_{j}   \label{a1}
\end{eqnarray}

\noindent
where $D$ is the total $x-$ derivative , $D^{-1}$ is the inverse operator
and all parameters are constants. In this work we consider
the system of equations (\ref{a0}) admitting recursion operators (\ref{a1})
of the irreducible (4,1)-type. A recursion operator will be called
irreducible if it is not possible to write it as ${\cal R}^{i}_{j}=
\tilde{R}^{i}_{k}\,\tilde{R}^{k}_{j}$
, where $\tilde{R}^{i}_{j}$ is the recursion operator of the type (2,1)
given by $\tilde{R}^{i}_{j}=b^{i}_{j}\,D^2+a^{i}_{jk}\,q^{k}+
c^{i}_{jk}\,q^{k}_{x}\,D^{-1}$.
KdV systems admitting reducible recursion operators of the type (4,1) belong
to the class studied recently \cite{MET1}.

Here we shall not give a systematic classification of this system for
all $N$. For $N=2$ we present the recursion operator and
 bi-Hamiltonian formulation of the DS system.
This system is a non-degenerate but contains
a nontrivial linear term $q^{i}_{x}$ . It is given in the following form

\begin{eqnarray}
u_{t} & = & -u_{xxx}+6 uu_{x}+6v_{x} ,\nonumber\\
v_{t} & = & 2 v_{xxx}-6uv_{x}. \label{a2}
\end{eqnarray}

\noindent

We find that the recursion operator ${\cal R}$ of this system is

\begin{equation}
{\cal R}=
\left(
\begin{array}{cc}
{\cal R}^{0}_{0}
 & {\cal R}^{0}_{1} \\
{\cal R}^{1}_{0}  &
{\cal R}^{1}_{1}
\end{array} \; \; \right)\;.
  \label{a3}
\end{equation}

\noindent

with

\begin{eqnarray}
{\cal R}^{0}_{0} & = & D^{4}- 8 uD^{2}
-12u_{x}D-8u_{xx}+16u^{2}
+16v+(-2\,u_{xxx}+12\,uu_{x}+ \nonumber \\
& & 12\,v_{x})D^{-1}+4u_{x}D^{-1}u,\nonumber\\
{\cal R}^{0}_{1} & = & -10D^{2}+8u+4u_{x}D^{-1},
\nonumber\\
{\cal R}^{1}_{0} & = & 10v_{x}D
+12v_{xx}+(4v_{xxx}-12uv_{x})D^{-1}+4v_{x}D^{-1}u,\nonumber\\
{\cal R}^{1}_{1} & = & -4D^{4}+16uD^{2}+8u_{x}D+16v+4v_{x}D^{-1}.
\nonumber
\end{eqnarray}

\noindent
Now it can be shown that this recursion operator
satisfies the hereditary property \cite{FUC1}. Furthermore it
admits the factorization \cite{FK1}
${\cal R}^{i}_{j}=(\theta_{2})^{ik}(\theta_{1}^{-1})_{kj}$
 where
\begin{equation}
\theta_{1}=
\: \left(
\begin{array}{cc}
0
 & {1 \over 2}D \\
{1 \over 2}D  &
{1 \over 4}D^{3}-{1 \over 2}(Du+uD)
\end{array} \; \; \right)\;.      \label{a4}
\end{equation}
and

\begin{equation}
\theta_{2}=
\: \left(
\begin{array}{cc}
\theta^{00}_{2}
 & \theta ^{01}_{2} \\
\theta^{10}_{2}  &
\theta ^{11}_{2}
\end{array} \; \; \right)\;.   \label{a5}
\end{equation}

\noindent
with

\begin{eqnarray}
\theta^{00}_{2} & = & -5D^{3}+4uD+2u_{x} \nonumber\\
\theta^{01}_{2} & = & -2D^{5}+4(D^{3}u+uD^3)+4Du_{x}D+4u_{x}D^2
+6Dv+2vD \nonumber\\
\theta^{10}_{2} & = & -2D^{5}+8uD^{3}+4u_{x}D^{2}+2Dv+6vD
 \nonumber\\
\theta^{11}_{2} & = & -D^{7}+4uD^{5}+2u_{x}D^{4}+2D^{4}(u_{x}
+2uD)+2vD^{3}+2D^{3}v \nonumber\\
& & -8uD^{2}(u_{x}+2uD)
 -4u_{x}D(u_{x}+2uD) -8v(u_{x}+2uD)-8uv_{x} \nonumber
 \end{eqnarray}

\noindent

Following the procedure of \cite{OLV} one can easily show that both
differential operators are skew-adjoint and satisfy the Jacobi
identities. Moreover they constitute a compatible pair.
Hence the system (\ref{a2}) can be
written in a bi-Hamiltonian form. Using  the compatible Hamiltonian
operators (\ref{a4}) and (\ref{a5}) we have

\begin{equation}
\left(
\begin{array}{c}
u \\ v
\end{array}\right)_{t}
=\theta_{1} \left(\begin{array}{c}
{\frac{\delta  {\cal H}_{1}}{\delta u}} \\ {\frac{\delta
{\cal H}_{1}}{\delta v}} \end{array} \right) \;
 =\theta_{2} \left(\begin{array}{c}
 {\frac{\delta {\cal H}_{0}}{\delta u}} \\ {\frac{\delta
 {\cal H}_{0}}{\delta v}} \end{array} \right) \;
\end{equation}

\noindent
associated with the Hamiltonian functionals

\begin{eqnarray}
{\cal H}_{0}[u,v] & = & \int({1 \over 2}u^{2}+v)dx,\nonumber\\
{\cal H}_{1}[u,v] & = & \int({1 \over 2}u_{xx}^{2}
-{5 \over 2}u^{2}u_{xx}+{5 \over 2}u^{4}\nonumber \\
&+&2u_{x}v_{x}+6(u^{2}v+v^{2}))dx.
\end{eqnarray}

\noindent
There thus exits a whole hierarchy of conservation laws and commuting
symmetries (flows) for the DS system.

We have a second commuting hierarchy originated from the
translational symmetry which may be formulated in a
bi-Hamiltonian structure

\begin{equation}
\left(
\begin{array}{c}
u \\ v
\end{array}\right)_{t}=
\left(
\begin{array}{c}
u \\ v
\end{array}\right)_{x}
=\theta_{1} \left(\begin{array}{c}
{\frac{\delta  {\hat{\cal H}}_{1}}{\delta u}} \\ {\frac{\delta
{\hat{\cal H}}_{1}}{\delta v}} \end{array} \right) \;
 =\theta_{2} \left(\begin{array}{c}
 {\frac{\delta {\hat{\cal H}}_{0}}{\delta u}} \\ {\frac{\delta
 {\hat{\cal H}}_{0}}{\delta v}} \end{array} \right) \;
\end{equation}

\noindent
where

\begin{equation}
{\hat{\cal H}}_{0}[u,v]  =  \int{1 \over 2}udx , \,\,\,\,\,
{\hat{\cal H}}_{1}[u,v]  =  \int({1 \over 2}u_{x}^{2}
+2uv+u^{3})dx .
\end{equation}

\noindent
are both conserved.

The interesting point here is that the linear term $v_{x}$ in this
system is nontrivial. In the case of the KdV systems admitting recursion
operators of type (2,1), the linear terms $\chi^{i}_{j}\,q^{j}_{x}$
are not essential in the study of the integrability of these systems.
This is based on the theorem given in ref.\cite{MET1}. It states that a KdV
system with  linear first derivative terms is integrable if and only if
its principle part (system without the linear first derivative terms) is
integrable. In the case of the KdV systems admitting recursion operators
of type (4,1) this theorem is not valid anymore .


We thank  Prof. V. Sokolov for his constructive comments.
This work is partially supported by the Scientific and Technical
Research Council of Turkey (TUBITAK) and by Turkish Academy of
Sciences (TUBA).


\begin{thebibliography}{99}

\bibitem{MET} M. G{\" u}rses and A. Karasu, Phys.Lett.{\bf A 214}, 21 (1996).
\bibitem{MET1} M. G{\" u}rses and A. Karasu,
J.Math.Phys.{\bf 39}, 2103 (1998).
\bibitem{ITO} M. Ito, Phys.Lett.,{\bf 91A} , 335 (1982).
\bibitem{FUC} B. Fuchssteiner, Prog.Theor.Phys.,{\bf 68}, 1082 (1982).
\bibitem{MA}  W.X. Ma and B. Fuchssteiner , Phys.Lett. {\bf A 213}, 49 (1996).
\bibitem{KUP} B.A. Kupershmidt , J.Phys. {\bf A 18}, L571(1985).
\bibitem{ANT} M. Antonowicz and A.P Fordy ,Physica {\bf D 28},345(1987).
\bibitem{ATH} C. Athorne and A.P. Fordy,  J.Phys. {\bf A 20}, 1377(1987).
\bibitem{LIU} Q. P. Liu , J. Math. Phys. {\bf 35}, 816 (1994).
\bibitem{OLV1} P.J. Olver and P.Rosenau, Phys. Rev. {\bf 53 E}, 1900 (1996).
\bibitem{FOK} A.S. Fokas and Q.M. Liu , Phys.Rev.Lett. {\bf 77}, 2347 (1996).
\bibitem{SVI} S.I. Svinolupov ,Theor.Mat.Fiz.,{\bf 87}, 391 (1991).
\bibitem{SV1} S.I. Svinolupov ,Functional Anal.Appl.{\bf 27}, 257 (1993).
\bibitem{OLV2} P.J. Olver and V.V.Sokolov, Commun. Math. Phys. {\bf 193},
               245 (1998).
\bibitem{HIR} R. Hirota and J. Satsuma, Phys.Lett.{\bf A 85}, 407(1981).
\bibitem{DOD} R. Dodd and A.P. Fordy,  Phys.Lett.{\bf A 89}, 168 (1982).
\bibitem{OEV} W. Oevel, Phys.Lett.{\bf A 94}, 404(1983).
\bibitem{AIY} R.N. Aiyer, Phys.Lett.{\bf A 93}, 368(1983).
\bibitem{WIL} G. Wilson, Phys.Lett.{\bf A 89}, 332(1982).
\bibitem{LEV} D. Levi, Phys.Lett.{\bf A 95}, 7(1983).
\bibitem{XIA} X. Geng and Y. Wu, J.Math.Phys.{\bf 38}, 3069 (1997).
\bibitem{MC} H. P.McKean,{\it Boussinesq's Equation as a Hamiltonian
 System} Topics in Functional Analysis.Adv.in Math.Suppl.Stud.
,{\bf 3}.pp.217-226 Academic Press,New York-London,1978.
\bibitem{ADL} M. Adler, Inven.Math.{\bf 50}, 219(1979).
\bibitem{OLV} P.J. Olver., {\it Applications of Lie Groups to
Differential Equations}, Second Edition, Graduate Texts in Mathematics,
Vol. 107, Springer-Verlag, New York, 1993.
\bibitem{DRI1} V.G. Drinfeld and V.V. Sokolov,{\it Proceedings of S.L
Sobolev Seminar, Novosibirsk},{\bf 2},5-9(1981) (in Russian).
\bibitem{DRI} V.G. Drinfeld and V.V. Sokolov,
  J.Sov.Math.{\bf 30}, 1975 (1985).
\bibitem{BOG} O.I. Bogoyavlenskii, Russian Math. Surveys,{\bf 45}, 1 (1990).
\bibitem{TIA} B. Tian and Y. Gao, Phys.Lett.{\bf A 208}, 193 (1995).
\bibitem{FUC1} B. Fuchssteiner, Prog.Theor.Phys.,{\bf 65}, 861 (1981).
\bibitem{FK1} B. Fuchssteiner and A. S. Fokas ,
 Physica {\bf 4D}, 47 (1981).
\end{thebibliography}
\end{document}